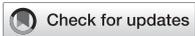

# Future accelerator projects: new physics at the energy frontier

Anadi Canepa[1]* and Monica D'Onofrio[2]*

[1]Fermi National Laboratory, Batavia, IL, United States, [2]Department of Physics, University of Liverpool, Liverpool, United Kingdom

High-energy colliders provide direct access to the energy frontier, allowing to search for new physics at scales as high as the machine's center-of-mass energy, perform precision measurements of the Standard Model (SM) parameters, including those related to the flavor sector, and determine the Higgs boson properties and their connection to electroweak symmetry breaking. Each proposed future collider option has its own specific science goals and capabilities, depending on the designed running energy (energies) amongst other parameters. In this paper, an overview of the discovery potential of future circular and linear colliders is presented. Results from searches for beyond the Standard Model (BSM) phenomena at proton–proton, proton–electron, electron–positron, and muon–antimuon colliders are summarized.

KEYWORDS

future, colliders, beyond the Standard Model, resonances, supersymmetry, hidden sector, dark matter, extended Higgs models

## 1 Introduction

Particle physics advances the fundamental description of "nature" at the smallest scales, leading and influencing global scientific efforts. The Large Hadron Collider (LHC) endeavor remains by far the major focus of the efforts and engagement of the particle physics community. During Run 1 and Run 2 of the LHC, the ATLAS and CMS experiments have produced first observations of fundamental processes, including the discovery of the Higgs boson in 2012 and the determination of its properties and couplings, and hundreds of Standard Model (SM) measurements and searches for new physics. Hints of lepton flavor universality violation (LFV) have been reported by the dedicated LHCb experiment, which also provides improved knowledge of quark mixing matrix parameters, such as the CKM angle, $\gamma$, and the discovery of many new hadronic states. Heavy ion studies are also pursued at the LHC by the specialized ALICE experiment, as well as by ATLAS and CMS. During the almost 12 years of LHC operation, there have been many experimental and theoretical advancements: it is remarkable that the precision of SM measurements and the sensitivity reach of new physics searches have exceeded, in some cases by far, the pre-LHC era expectations.

The experimental success of the LHC is certainly a result of the excellent performance of the detectors and the accelerator complex, and its high luminosity upgrade, the HL-LHC, will maximize its potential. Scheduled to run until 2038–2040, the HL-LHC program will allow the general purpose detectors to collect an integrated luminosity of 3 ab$^{-1}$ of proton–proton collisions at a center-of-mass energy (or $\sqrt{s}$) of 14 TeV. The determination of the Higgs boson properties, and their connection to electroweak symmetry breaking (EWSB), is one of the primary science goals of ATLAS and CMS [1]. Outstanding opportunities will emerge for measurements of fundamental importance, such as the first direct constraints on the Higgs trilinear self-coupling and its natural width. SM processes and parameters, including those related to the flavor sector, will be tested by performing measurements with unprecedented precision [2], such as the production of pairs or triplets of EW gauge bosons, the effective weak mixing angle, and the masses of the top-quark and W-boson. On the latter, it is noticeable that the Large Hadron electron Collider (LHeC) [3], designed to collide an intense electron beam produced by an energy recovery linac with the HL-LHC proton beam, would allow to reduce the uncertainty on this fundamental parameter to 2 MeV. The search for physics beyond the SM (BSM) will remain the main driver of the exploration program at the HL-LHC [4]. In most BSM scenarios, the HL-LHC will increase the present reach in mass and coupling by at least 20%–50% and will allow searches for, among others, additional Higgs bosons in EWSB scenarios, new resonances, candidates for dark matter (DM), and hidden sectors to be performed. The LHCb program will continue to focus on heavy-flavor physics throughout the HL-LHC phase [5], possibly confirming the anomalies in the lepton-flavor sector and, as such, establishing LFV and opening a new era of discoveries. LHCb will also search for feebly interacting new particles arising in hidden sectors, complementing ATLAS, CMS, and other beyond-collider experiments in these searches.

The conclusion of the European Strategy for Particle Physics update (ESPPU) and the ongoing Snowmass and Particle Physics Project Prioritization Panel (P5) process presents an opportunity to revisit the landscape of future accelerators beyond the HL-LHC. Several lepton and hadron collider options have been considered, each with merits and challenges, and emphasis has been given on attainable physics targets as well as technological requirements and drivers.

This review focuses on the proposed accelerators' potential for discovering physics beyond the SM, and it is primarily based on feasibility studies reported in the literature at the time of the ESPPU process. If they were available at the time of writing, updates on searches and measurements submitted through the Snowmass process have been included as well. Considerations regarding the status of the proposed machines are also presented.

## 2 Overview of proposed collider options

Extensive activities have been carried out worldwide to assess the future of collider experiments beyond the HL-LHC. A summary of the accelerator-based projects proposed by the community in recent years and considered in this paper is presented in Table 1.

Electron–positron colliders (linear or circular) with $O$(100 GeV) center-of-mass energy such as the Future Circular Collider $e^+e^-$, FCC-ee [6], the International Linear Collider, ILC [64], and the Compact Linear Collider, CLIC [7] represent primarily a Higgs factory providing unprecedented precision measurements of the Higgs boson properties. At the FCC-ee, this would come in combination with improvements in the knowledge of the SM couplings from the planned Tera-Z facility, also enabling an interesting flavor physics program. At linear colliders, a significant advance in precision is also expected thanks to the available beam polarizations. The $e^+e^-$ machines are also discovery machines, with high center-of-mass energy options such as CLIC$_{3000}$, extending the sensitivity to high-mass phenomena and, in some areas, yielding a reach comparable to high-energy $pp$ colliders. For similar center-of-mass energies and integrated luminosities, the CepC [8] physics program and potential are comparable to those of the FCC-ee, chosen here as a representative case.

The future 100 TeV center-of-mass energy proton–proton collider at the FCC, FCC-hh [9], expected to run after the completion of the FCC-ee stages, offers several unique possibilities for a breakthrough in particle physics. Aiming to collect integrated luminosities up to 20 ab$^{-1}$ per experiment in 10 years of operation, FCC-hh will allow measurements of the Higgs self-coupling at the few percent level, measurements of quartic Higgs self-coupling, and direct searches for new physics at the highest energy, with the possibility to extend by an order of magnitude the LHC sensitivity above the EWSB scale. Precise differential cross-section measurements for high-transverse momentum Higgs boson production and measurements of rare processes will be possible due to the large datasets. With the addition of an energy recovery electron linac of 60 GeV, electron–proton interactions could be explored [10], providing additional inputs to achieve the ultimate Higgs physics precision at the FCC-hh, QCD precision measurements, and searches for new physics. The FCC hadron complex would also allow for a dedicated heavy ion program, with lead–lead and proton–lead, and electron–lead collisions possible at the FCC-hh and FCC-eh, respectively.[1] An alternative idea for a $pp$ collider considers the possibility of increasing the energy of the LHC machine up to 27–30 TeV, turning the current accelerator into a high-energy machine (HE-LHC [12]). This would allow utilizing the current tunnel and the entire CERN infrastructure with future magnet technologies to collect large datasets at $\sqrt{s}$ at least two times the one of the HL-LHC.

TABLE 1 Collider options whose physics case is discussed in this paper. Configurations and parameters are those proposed at the time of writing (see Section 10 for more details). In case multiple stages are foreseen for a specific project, all center-of-mass energies and benchmark luminosity values are reported. In the text and figures, the ILC stages are indicated as ILC$_{250}$, ILC$_{500}$, and ILC$_{1000}$; CLIC's three phases are referred to as CLIC$_{380}$, CLIC$_{1500}$, and CLIC$_{3000}$. A similar nomenclature is adopted for FCC-ee and the muon collider (MuC) where relevant. The integrated luminosity is per interaction point (IP), except for the FCC-ee and the CepC, where it corresponds to two IPs.

| Collider (type) | $\sqrt{s}$ (GeV) [$\mathcal{L}_{int}$ (ab$^{-1}$), duration (years)] |
|---|---|
| HE-LHC (circular, $pp$) | $27 \times 10^3$ [15, 20] |
| ILC (linear, $e^+e^-$) | 91 [0.1, 1.5]; 250 [2, 11]; 350 [0.2, 0.75]; 500 [4, 9] |
| CLIC (linear, $e^+e^-$) | 380 [1.0, 8]; $1.5 \times 10^3$ [2.5, 7]; $3 \times 10^3$ [5, 8] |
| FCC-ee (circular, $e^+e^-$) | 88–94 [150, 4]; s-channel h [20, 3]; 157–163 [10, 2]; 240 [5, 2]; 340–365 [1.7, 5] |
| FCC-hh (circular, $pp$) | $100 \times 10^3$ [20–30, 25] |
| FCC-eh (circular plus ERL, $ep$) | $3.5 \times 10^3$ [3, 25] |
| MuC (circular, $\mu^+\mu^-$) | 3 TeV [1, 5]; 10 TeV [10, 5]; 10 TeV [20, 5] |
| CepC (circular, $e^+e^-$) | 91 [16, 2]; 160 [2.6, 1]; 240 [5.6, 7]; 360 [-, -] |

A $\mu^+\mu^-$ collider, MuC [13], could give the opportunity to achieve a multi-TeV energy domain beyond the reach of the $e^+e^-$ colliders and within a much shorter circular tunnel than for a $pp$ collider. The picture emerging from studies of the past years is that a 10 TeV muon collider could combine the advantages of $pp$ and $e^+e^-$ colliders due to the large $\sqrt{s}$ available for direct exploration and to the achievable accuracy for precise measurements of the SM. By exploiting the copious rate of vector–boson fusion and vector–boson scattering processes, a MuC provides the opportunity to probe details of the EWSB mechanism. Muon-philic new physics scenarios, possibly explaining the $g − 2$ [14] and $B$-physics anomalies [15], are additional natural targets. Because a muon production and cooling complex could be used at all energies, and muon acceleration proceeds through a sequence of rings, a $\mu^+\mu^-$ collider can be built in stages, with 3 TeV center-of-mass energy foreseen as the first stage.

Future collider concepts [16] not explicitly listed previously are also being considered within the Snowmass/P5 process. Among those, the $C^3$ linear collider project [17] could fit on the Fermilab site and would have a similar potential to that of the aforementioned $O$(100 GeV) machines, with a starting center-of-mass energy of 250 GeV to be potentially increased to 550 GeV and to 3 TeV by extending the accelerator's length. Other options based on high-gradient superconducting radio frequency (SRF) technology, such as the compact SRF Higgs-Energy LEptoN (HELEN) linear collider, are also being investigated [18]. In terms of circular machines, a 16-km circumference circular $e^+e^-$ collider with center-of-mass energy between 90 GeV and 240 GeV is being examined with Fermilab as a potential site, as well as a possible proton–proton collider with center-of-mass energy between 24 and 27 TeV located in the same tunnel, with a reach similar to that of a HE-LHC. Finally, a Super proton–proton Collider (SppC [19]) is proposed as a machine located in China, running after the CepC, and using the same tunnel complex and infrastructure in a multi-staged approach similar to that envisaged for the CERN FCC. More details are given in Section 10.

The planning spans a 30-year horizon, as major accelerator-based projects require developments on that timescale. Comparing the physics potentials, the required technology and prospects for its availability, and the cost-to-benefit ratio of the proposed machines is extremely challenging. Each collider program, to varying degrees and dependent in part on the center-of-mass energy considered, gives good coverage of almost all fundamental physics questions. They also have unique synergies with the neutrino and precision frontiers, as well as with astrophysics and cosmological investigations ongoing or planned during the next decades. An overview of those complementarities is, however, beyond the scope of this paper.

# 3 Searches at colliders: physics landscape

The SM has been proven very successful in describing elementary particles and their interactions. It has been validated extensively through precision experiments, and the discovery of the Higgs boson has certainly been a major milestone in this respect. However, there are a number of shortcomings and several open questions that the SM fails to answer. Severe fine-tuned cancellations of large quantum corrections are required to obtain a Higgs boson mass close to the EW scale, leading to the so-called hierarchy problem. The SM does not incorporate gravity as described by general relativity, or account for the accelerating expansion of the universe. It does not contain any viable dark matter particle and fails to explain in full baryon asymmetry, neutrino oscillations, and non-zero neutrino masses. As such, a plethora of theories beyond the SM have been developed in the past decades, and the search for them is at the core of the particle physics community's experimental activities. While formulating an exhaustive and complete classification of all existing BSM models is not possible, it is evident that the exploration of the unknown is one of the main drivers of all future colliders:

- Important goals of future colliders include searches for the existence of new gauge or space-time symmetries and tests of theories containing multi-TeV resonances. Mostly related to the dynamics of EWSB, vector resonances, leptoquarks, and contact interactions are among the BSM theories considered in this paper. Direct searches for heavy new particles can be complemented by precision studies of SM observables, and deviations from predictions would be an indirect but powerful way to provide evidence of new physics.
- Several new physics models focus on the nature of the Higgs boson, either considering the possibility that it is a composite state or that it belongs to an extended sector with new scalar particles, where one closely resembles the SM Higgs boson. For the latter, various models with different Higgs representations have been proposed. Among those receiving the most attention for future collider studies is the extension of the SM scalar potential by a singlet massive scalar field that can change the nature of the EW phase transition. Another common set of extended Higgs sector models searched for is characterized by the addition of a second SU(2) doublet, which naturally appears in supersymmetric extensions of the

---

[1] We note that the nuclear physics community is also pursuing the Electron-Ion Collider (EIC) project, the first-ever collider of polarized electrons with nuclei or polarized protons. The target center-of-mass energy (20–140 GeV) is substantially smaller than that achieved at HERA, but the target luminosity is foreseen to be 1,000 times higher. Considerable synergies with accelerator particle physics projects in terms of detector technology and physics potential are expected. For a detailed report on the EIC project, see Ref. [11].

- Higgs sector or in models with a non-minimal pattern of symmetry breaking.
- Supersymmetry (SUSY) certainly remains one of the most plausible beyond the SM scenarios, as it provides the only known dynamical solution to the Higgs naturalness problem that can be extrapolated up to very high energies. SUSY gives an excellent potential candidate for DM as well as a framework for gauge coupling unification and possibly reconciliation of gravity and other forces. As such, it is the focus of multiple studies from various facilities both in the strong and electroweak sectors.
- Cosmological data suggest that DM particles could have masses in the range from multi-keV to approximately 100 TeV and couplings to SM particles of comparable or weaker strength than EW interactions. High-energy colliders could produce DM particles within this mass range in controlled conditions and, as such, complement experiments and observations from astroparticle physics experiments. A typical DM thermal relic studied at colliders is a weakly interacting massive particle, referred to as WIMP. Several DM models predict the presence of mediator particles, whose exchange may be responsible for the annihilation processes that determine the DM particle abundance and can be directly searched at colliders. If the DM particle is lighter than $m_h/2$ and it is coupled to the Higgs, a compelling exploration channel is an invisible Higgs decay. Of particular interest are the cases of spin-1/2 particles transforming as doublets or triplets under SU(2) symmetry.
- An alternative possibility for new physics is that particles responsible for the still unexplained phenomena have not been detected because they interact too feebly with SM particles. These particles could belong to an entirely new sector, the so-called hidden or dark sector. While masses and interactions of particles in the dark sector are largely unknown, the mass range between the keV and tens of GeV appears interesting, both theoretically and experimentally.
- Heavy new physics can induce, through the exchange of virtual particles, processes that are extremely rare in the SM, such as flavor-changing neutral current (FCNC) effects in the top-quark sector. The expected intensity of some of the future lepton collider proposals at critical production thresholds will allow improvements in sensitivity. Hadron colliders at very high luminosities and muon colliders at multi-TeV center-of-mass energy are also complementary when looking for rare processes.
- In the absence of evidence for new physics at low energy and assuming that BSM is realized at a scale $\Lambda$ much larger than the collider $\sqrt{s}$, the effective field theory (EFT) formalism is adopted as a framework to study BSM physics with a model-independent approach. Several EFT representations exist, and a subset of those is considered in this review.

This program is continuously evolving and broadening in response to results from the current LHC, the HL-LHC, and other ongoing and future non-collider experiments. Beyond-collider projects in construction, planned, or proposed to further exploit the LHC accelerator complex will significantly boost the discovery potential in the next two decades, offering complementarities and synergies in the quest for new physics. Among those, FASER [20] and SND@LHC [21] will start operations during the Run 3 of the LHC; others like MATHUSLA [22], CODEX-b [23], MilliQan [24], and the LHeC [25] are foreseen to operate in parallel to the HL-LHC. An extensive proposal on the Forward Physics Facility at CERN has been presented recently [26], while Ref. [27] presents detectors for fixed-target experiments and beam-dump experiments at the ILC complex.

A subset of representative prospective results is reported in the rest of this review. They originate from hundreds of studies of varying degrees of sophistication carried out over several years. Detailed or fast simulations are used in some cases, whilst simple detector parameterizations, direct extrapolations of results from existing data, or even simple rescaling are performed in other cases. The reader is referred to the original publications for details on the analyses and on the approach and hypotheses made.

## 4 New resonances

On-shell resonances decaying into visible SM particles are a distinct signature of several BSM theories, ranging from new models of EWSB to extensions of the SM gauge group. Classic scenarios include singly produced resonances with integer spin or pair-produced heavy resonances. Direct access requires the center-of-mass energy of the collider to be large enough to produce them. Performances can then be evaluated considering the reach in mass, or the reach in mass vs. coupling, with the results depending on the assumptions on the couplings of the new particle to quarks and leptons. If the colliders' center-of-mass energies are below the mass of the new hypothetical resonance, indirect access can be achieved by studying deviations in SM observables.

One of the most widely used benchmark scenarios predicts a new high-mass vector (spin-1) boson, the $Z'$. Examples [28] are the sequential SM (SSM), $B-L$, and $E_6$ $Z'$ models, as well as models of little Higgs or extra dimensions. The primary discovery mode for a $Z'$ at hadron colliders is the Drell–Yan production of a dilepton resonance, but hadronic final states are also widely studied. The mass reach is typically in the $(0.3 - 0.5)\sqrt{s}$ range, given sufficient statistics. FCC-hh [29] could discover a SSM $Z'$ with a mass of up to 43 TeV if it decays into an electron or muon pair, assuming 30 ab$^{-1}$ of luminosity. Masses between 20 and 30 TeV could be reached if decays to $\tau^+\tau^-$ or to $t\bar{t}$ are instead considered. A multi-TeV MuC could become competitive in accessing directly a $Z'$, especially in the case of muon-philic models [30], where the new vector boson dominantly couples to $\mu^+\mu^-$, e.g., via left-handed currents.

Given the current mass limits from the LHC experiments, a direct observation of these new resonances is not expected at the currently planned linear and circular $e^+e^-$ accelerators for most of the scenarios considered in the literature. Nonetheless, the presence of high-mass resonances can be inferred indirectly using an EFT approach to describe BSM virtual effects. In the EPPSU studies, a benchmark model dubbed "Y-Universal $Z'$" has been used for a quantitative assessment of the potential of future colliders to search for new gauge bosons, directly and indirectly [31]. Figure 1 (left) shows the 95% confidence level (CL) limits in mass vs. coupling at various colliders. The model assumes the same couplings, $g_{Z'}$, to quarks and leptons, and it was chosen because it allows for a fair comparison between hadron and lepton colliders. The direct constraints from FCC-hh are the most stringent at low $g_{Z'}$, while indirect reaches of both $e^+e^-$ and $pp$ colliders become superior at high $g_{Z'}$. At lepton colliders, an EFT framework allows to achieve sensitivity thanks to the EW precision measurements of the oblique parameter $S$ [33]. At hadron colliders, Drell–Yan predictions are also sensitive to the ratio between $g_{Z'}$ and the $Z'$ mass. As such, very precise parton distribution function (PDF) fits obtained using $ep$ collisions would further improve the sensitivity [3]. Finally, the muon collider reach, not reported in this figure, is estimated to be similar to that of CLIC$_{3000}$ for $\sqrt{s}$ = 3 TeV and exceeding it proportionally to the increase in center-of-mass energy [34].

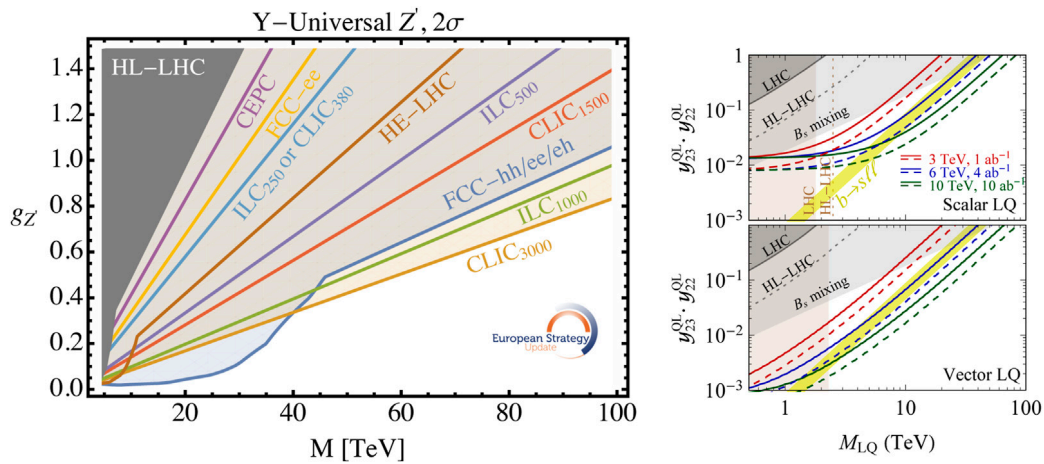

FIGURE 1
Left: Exclusion reach of different colliders on the Y-Universal $Z'$ model parameters [31]. Right: Muon collider sensitivity contours at 95% CL for scalar (upper panel) and vector (lower panel) LQ models via the process $\mu^+\mu^- \to bs$, where $y^{QL}_{ij}$ indicates the coupling between the $i$-generation lepton and the $j$-generation quark. For the various setups considered, see Ref. [32].

Expected sensitivity on the production and decay of spin-0 and -2 particles decaying into several different SM final states has also been studied at high-energy lepton and hadron colliders. Models considered include, among others, resonant double-Higgs production and heavy scalar singlets that could mix with the Higgs boson, i.e., see Refs [35–37].

Leptoquark (LQ) models, alongside $Z'$ models, have gained considerable renewed interest in recent years as they can give rise to lepton universality violating decays of heavy mesons at the tree level, provided that couplings are generation-dependent and they couple to the second and third generations of quarks. LQs are hypothetical particles that carry both baryon and lepton quantum numbers. They are color-triplets and carry fractional electric charge. The spin of a LQ state is either 0 (scalar) or 1 (vector). Models predicting a rather light LQ coupled predominantly to the third generation are a natural target for hadron colliders where scalar or vector LQs are pair-produced via strong interaction and results [31] are independent of the coupling to the lepton quark current. If discovered, FCC-eh could contribute to their characterization, assuming that the coupling to the first-generation quark is non-negligible and can be produced as an $s$-channel resonance [3]. Muon colliders have the best sensitivity for a LQ model via $\mu^+\mu^- \to bs$. With a few to 10 TeV center-of-mass energy and predicted luminosities of 1–10 ab$^{-1}$, a MuC could cover the entire parameter space that explains the flavor anomalies for both scalar and vector LQ. Results are shown in Figure 1 (right), from Ref. [32].

If new particles arising in BSM theories are much heavier than the energy reach for on-shell production even at future colliders, their existence can still be formalized through contact interactions (CIs). An effective four-fermion CI could represent the exchange of a virtual heavy particle, such as an LQ, a $Z'$, or elementary constituents of quarks and leptons in composite models. The effective CI scale represents the typical mass scale of the new particles, and the experimental sensitivity increases significantly with $\sqrt{s}$. Lepton colliders are powerful in testing the neutral-current case, owing to the precision that can be achieved in analyses of di-fermion final states with suitable statistics. Linear colliders can also exploit different longitudinal polarizations of the two beams. Hadron colliders have excellent sensitivity up to their $\sqrt{s}$ via Drell–Yan production for both neutral and charged currents. The highest reach as reported in the EPSSU studies [31] is up to 120 TeV (CLIC$_{3000}$). The so-called two-fermion/two-boson CIs are also phenomenologically relevant for BSM theories of EWSB because they describe new physics effects in the interaction between the gauge and Higgs sectors. In this case, estimated reaches [31] are, at best, 30–35 TeV. Precision differential measurements of the $ZH$ production provide the lead sensitivity for lepton colliders. Hadron colliders' sensitivity mostly comes from precision measurements of SM diboson production observables, as used in the FCC-hh studies. Additional studies on CIs related to new physics models possibly contributing to the muon $g-2$ and to high-energy scattering processes have also been carried out at the muon collider [30].

## 5 Composite Higgs and extended sectors

The role of the Higgs boson could be even more complex than that known so far in the SM formulation, and hence it is logical to also question its nature and whether or not it is a point-like particle. Composite Higgs models (CHMs) predict that the Higgs is not an elementary particle and that new particles might arise as excitations of the composite Higgs, with mass possibly at the $O(TeV)$ scale. The foundation of composite Higgs models is that the Higgs emerges as a bound state of a new strongly interacting confining composite sector, analogous to QCD, but with a much higher confinement scale. The Higgs, similarly to the pions in QCD, emerges as a Goldstone boson associated with a spontaneously broken global symmetry of the composite sector. The phenomenology of CHMs is mainly controlled by two parameters: the mass scale $m_*$, which controls the mass of the new resonances, and the coupling $g_*$, representing the interaction strength among particles originating from the composite sector. EFT operators that describe the indirect effects of Higgs compositeness at low energy are then defined, and their scale is set by $g_*$.

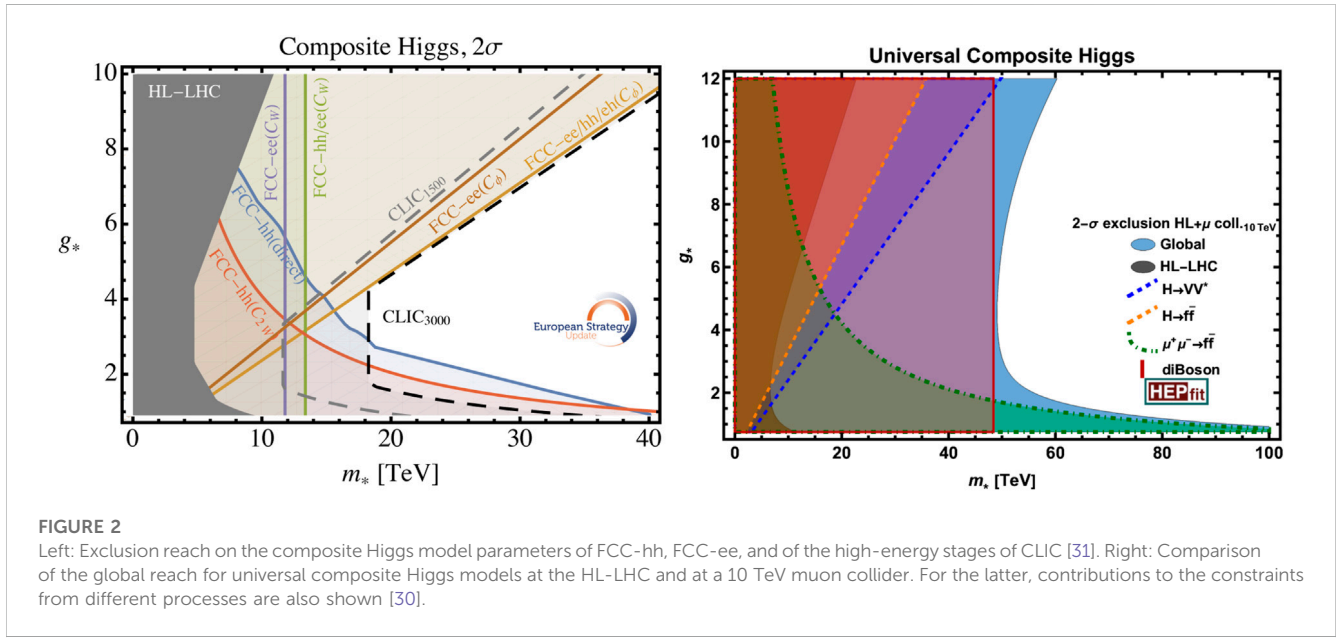

FIGURE 2
Left: Exclusion reach on the composite Higgs model parameters of FCC-hh, FCC-ee, and of the high-energy stages of CLIC [31]. Right: Comparison of the global reach for universal composite Higgs models at the HL-LHC and at a 10 TeV muon collider. For the latter, contributions to the constraints from different processes are also shown [30].

Figure 2 (left) shows the exclusion reach on $m_*$ and $g_*$ for FCC-hh, FCC-ee, and the high-energy stages of CLIC. Contours for the reach of HE-LHC, ILC, CepC, and $CLIC_{380}$ are also available in Ref. [31]. The 95% CL exclusion contours of each collider project arise from effects on coupling measurements [38] of the Higgs boson related to its possible composite nature and, for the FCC-hh and the HL-LHC, also from direct searches for an EW triplet $\rho$ vector resonance in dilepton and diboson final states [43]. Figure 2 (right) presents the exclusion reach for the 10 TeV stage of the muon collider. The reach of HL-LHC is also reported and statistically combined in the global result [30]. Other curves denote the contributions to the constraints from different processes, including that of searches for composite Higgs fermionic top partners.

Theories predicting an extended Higgs sector acquired significant prominence in the experimental programs of collider experiments, with searches targeting a broad spectrum of models. In minimal scenarios, the Higgs sector is augmented by a singlet massive scalar field which, e.g., can mix with the SM Higgs boson with a mixing parameter $\gamma$. The presence of the singlet can either modify the SM Higgs boson properties or be detected as single production of the massive particle associated with the field,

S, which subsequently decays into SM particles. Figure 3 (left) summarizes the reach [31], in the mass-$\sin^2\gamma$ space, of direct searches and indirect constraints derived from the Higgs boson couplings measurements (horizontal lines). Among the indirect searches, those performed at $CLIC_{3000}$ are the most sensitive searches and allow to probe mixing angles for values as low as $\sin^2\gamma \sim 10^{-3}$ for any value of $m_S$. In contrast, the reach of direct searches depends on the singlet's mass. The muon collider at $\sqrt{s}$ = 14 TeV explores masses as high as 9 TeV while extending the sensitivity to $\sin^2\gamma$ by almost one order of magnitude with respect to the best indirect constraint from $e^+e^-$ colliders. Thanks to the larger center-of-mass energy, the FCC-hh is sensitive to higher masses but yields a more limited reach in the mixing values.

Under the no-mixing assumption, the singlet-associated particle S would be stable and thus searched for in events with significant missing transverse momentum. The best sensitivity is yielded by indirect searches based on the precision measurement of the SM Higgs couplings at $CLIC_{3000}$, probing masses between 50 and 350 GeV and $\lambda_{HS}$ between 0.1 and 1, where $\lambda_{HS}$ is the coupling term in the potential $V \sim \lambda_{HS}|H|^2 S^2$ [31]. Experiments at the FCC-hh achieve a similar sensitivity through direct searches for the pair production of S. It is interesting to note that this region of phase space is compatible with a strong first-order EW phase transition, demonstrating that colliders have the potential to test models predicting the baryon asymmetry in the universe and gravitational waves. As such, the energy frontier complements the program at cosmology experiments, like the future gravitational wave experiment LISA. Another example is presented in Ref. [30], where the reach of a 3 TeV muon collider is compared to that of LISA.

More complex scenarios extending the Higgs sector by a new $SU$[2] doublet, e.g., supersymmetry or more generically type-II two-Higgs doublet models, predict the existence of two CP-even scalars, $h$ and $H$, one CP-odd scalar, $A$, and a charged scalar, $H^\pm$. This rich phenomenology leads to a variety of probes at future machines. As an example, Figure 3 (right) shows the constraints on $m_A$ as a function of $\tan\beta$, the ratio of the vacuum expectation value for the two Higgs doublets. While precision measurements of the Higgs couplings to third-generation fermions offer sensitivity to models at low values of $m_A$ (~1 TeV), the FCC-hh probes the existence of the new states for masses as high as 10–20 TeV for any values of $\tan\beta$. Assuming efficient signal detection and background suppression, the sensitivity to the pair production of new states at lepton colliders goes up to $m \leq \sqrt{s}/2$. Measurements of flavor physics observables can also lead to constraints on the type-II two-Higgs doublet models (i.e., see Ref. [39]).

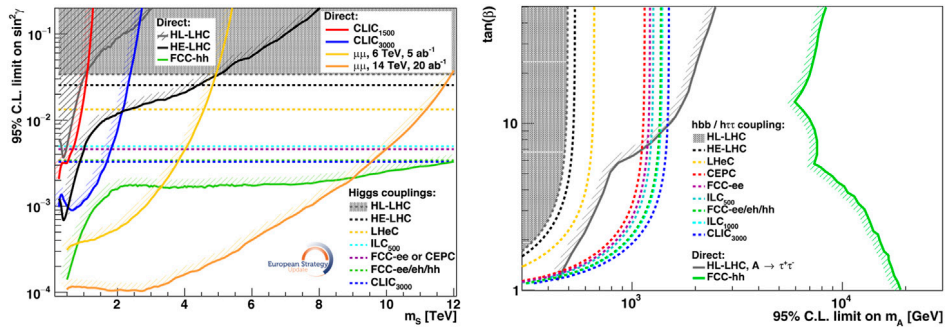

FIGURE 3
Exclusion reach obtained from the precision measurements of the Higgs boson couplings and from direct searches for new states. Left: Sensitivity in the plane $\sin^2\gamma - m_S$, where $\gamma$ and $m_S$ are the mixing angle and the mass of the singlet mixing with the SM Higgs boson, respectively. Right: Sensitivity in the plane $\tan\beta - m_A$, where $\tan\beta$ is the ratio of the vacuum expectation value for the two Higgs doublets and $m_A$ is the mass of the CP-odd scalar Higgs boson, in a type-II two-Higgs doublet model [31].

# 6 Supersymmetry

The phenomenology of SUSY is mostly driven by its breaking mechanism and breaking scale, which define the SUSY particle masses, the mass hierarchy, the field contents of physical particles, and thus their cross sections and decay modes. In addition, signal topologies crucially depend on whether $R$-parity, defined as $R = (-1)^{3(B-L)+2S}$, where $B$ and $L$ are baryon and lepton numbers, respectively, and $S$ is the spin, is conserved or violated.

Indirect constraints from flavor physics experiments, high-precision electroweak observables, including the discovery of the 125 GeV Higgs boson, and astrophysical data impose strong constraints on the allowed SUSY parameter space. Still, SUSY can be the key to understand Higgs naturalness, and in $R$-parity conserving scenarios, the lightest supersymmetric particle (LSP) is an excellent candidate for DM. These are certainly strong motivations to search for colored SUSY particles, squarks, and gluinos, for EW gauginos and Higgsinos that mix into neutralino and chargino mass states, collectively referred to as electroweakinos (EWkinos, $\chi$), and for the superpartner of charged and neutral leptons, the sleptons.

Squarks and gluinos are produced via the strong interaction and have the highest cross sections at hadron colliders. Scalar partners of the left-handed and right-handed chiral components of the bottom-quark and top-quark mix to form mass eigenstates for which the bottom and top squarks are defined as the lighter of the two ($\tilde{b}_1$ and $\tilde{t}_1$, respectively) and might be significantly lighter than the other squarks and the gluinos. EWkinos cross sections depend on mixing parameters and are typically much smaller than those of colored superpartners at hadron colliders. For this reason, the EW sector remains more difficult to test at hadron machines, and searches at $e^+e^-$ colliders would complement the SUSY parameter space coverage. Similar considerations can be made for slepton pair production, as cross sections can be up to two orders of magnitude smaller than those for EWino pair production.

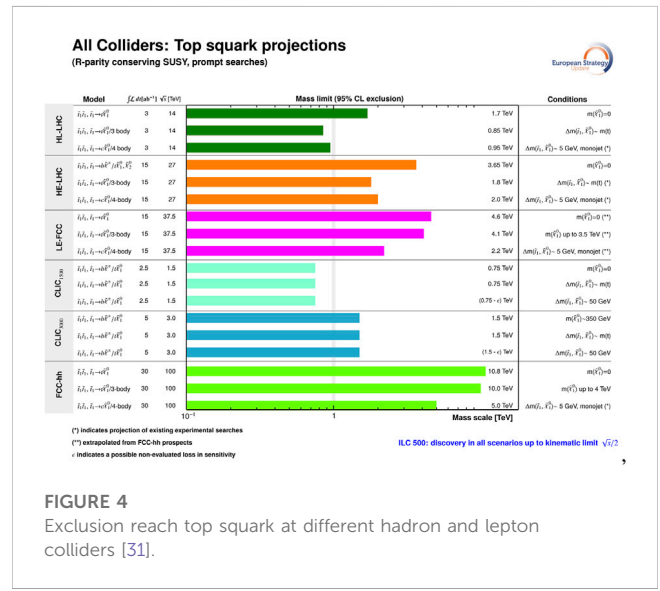

FIGURE 4
Exclusion reach top squark at different hadron and lepton colliders [31].

Prospects for SUSY searches are presented in terms of mass exclusion limits at 95% CL. The corresponding definitive observation with a significance of $5\sigma$ would be 5%–10% lower depending on the process. High-energy $pp$ colliders provide the most stringent bounds on first- and second-generation squarks and gluinos. In $R$-parity conserving scenarios, gluino (squark) masses up to 17 [9] TeV could be reached by the FCC-hh exploiting the typical multijet plus missing transverse momentum SUSY signature for a massless LSP, while monojet-like analyses, where the SUSY particles recoil against an initial state radiation (ISR) jet, are most effective for compressed scenarios. Lepton colliders are ineffective in the searches for gluinos, which are neutral with respect to the EW interaction, while current limits on first- and second-generation squark masses make the results not competitive. Similar conclusions can be drawn for top-squark pair-production searches if the preferred decay is $\tilde{t}_1 \to t\tilde{\chi}_1^0$ and $\Delta m(\tilde{t}, \tilde{\chi}_1^0) \gg m_t$, where $\tilde{\chi}_1^0$ is the lightest neutralino and $m_t$ is the top-quark mass. On the other hand, for small mass splitting, the sensitivity of $pp$ colliders significantly degrades so that high-energy lepton colliders, e.g., CLIC$_{3000}$ and MuC at 3–10 TeV, become competitive [34]. Their stop mass reach is close to $\sqrt{s}/2$ even for low $\Delta m(\tilde{t}, \tilde{\chi}_1^0)$, although a loss in acceptance and efficiency could be expected for mass differences of the order of 50 GeV. The exclusion limits are summarized in Figure 4; see [31] and references therein for details on the assumptions.

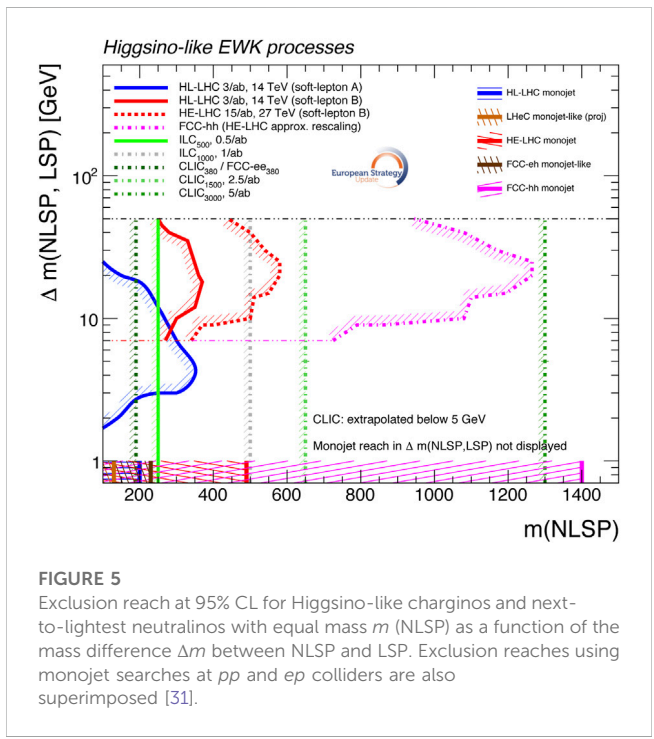

**FIGURE 5**
Exclusion reach at 95% CL for Higgsino-like charginos and next-to-lightest neutralinos with equal mass $m$ (NLSP) as a function of the mass difference $\Delta m$ between NLSP and LSP. Exclusion reaches using monojet searches at $pp$ and $ep$ colliders are also superimposed [31].

At hadron colliders, the largest production rates for EWkinos are obtained when the lightest chargino ($\tilde{\chi}_1^\pm$) and next-to-lightest neutralino ($\tilde{\chi}_2^0$) are Wino-like, forming an approximately mass degenerate SU(2) triplet referred to as the next-LSP (NLSP). Exclusion reaches for $R$-parity conserving Wino-like scenarios show that NLSP masses up to 3.3 TeV can be excluded at FCC-hh for a massless LSP, to be compared to a sensitivity of up to EWkino masses equal to $\sqrt{s}/2$ for high-energy $e^+e^-$ and $\mu^+\mu^-$ colliders even for $\Delta m(\tilde{\chi}_1^\pm, \chi_1^0)$ as low as 1 GeV, with no loss in acceptance. If the Higgsino mass is much smaller than the gaugino masses, the production rates are smaller, $\tilde{\chi}_{1,2}^0$ and $\tilde{\chi}_1^\pm$ form an approximately mass degenerate SU(2) doublet, and the EWkino spectrum is compressed. Feasibility studies and projections are summarized in Figure 5 (for details, see [31] and references therein). The sensitivity of lepton colliders depends only weakly on the nature of the LSP as cross sections are less dependent on the choice of mixing parameters. The high-energy lepton colliders allow a reach close to the pair production threshold, approximately

1.3 TeV for CLIC$_{3000}$, with the mass splitting down to approximately 0.5 GeV, and it is similar (not shown) for a MuC of 3 TeV center-of-mass energy. Beam polarization effects might also play a crucial role in searches for EWK SUSY at linear colliders [40, 41]. Analyses exploiting ISR jets and/or soft-momentum leptons show good prospects at hadron colliders in the case of Higgsino-like scenarios: $\tilde{\chi}_1^\pm, \tilde{\chi}_2^0$ masses up to approximately 600 GeV can be probed at the HE-LHC for mass splittings $\Delta m \equiv \Delta m(\tilde{\chi}_2^0, \tilde{\chi}_1^0) \approx \Delta m(\tilde{\chi}_1^\pm, \tilde{\chi}_1^0)$ between 7 and 50 GeV. FCC-hh projections show expected 95% CL limits up to 1.3 TeV, also depending on $\Delta m$, with monojet searches possibly complementing the reach for very compressed scenarios. Prospects for $ep$ colliders (LHeC and FCC-eh) performed using monojet-like signatures are also shown. Finally, if the lightest neutralino is either pure Higgsino or Wino, EWinos' mass splittings are theoretically calculated to be approximately 340 MeV and 160 MeV, respectively. In these cases, taking advantage of the long lifetime of the charginos, searches for disappearing charged tracks can be performed at hadron and electron–hadron colliders. Analyses exploiting displaced decays of the charged SUSY state have also been studied for lepton colliders. Results can be interpreted in the context of generic DM models and are reported in Section 7.

Significant sensitivity to sleptons is expected at future accelerators. High-mass selectrons, smuons, and staus are best accessed by hadron colliders for large mass splitting between the slepton and the LSP masses, with limits up to or in excess of 5 TeV for the FCC-hh [42], significantly depending on the assumptions of slepton handedness and mass degeneracy. Dedicated searches for staus, on the other hand, might be particularly challenging at $pp$ colliders because of the potentially high rate of misidentified tau leptons. The HE-LHC would provide sensitivity of up to 1.1 TeV, and an additional three-fold increase is expected for the FCC-hh [31]. Prospect studies at linear lepton colliders [43–45] show excellent expected sensitivity to slepton masses up to close to $\sqrt{s}/2$ and good potential for characterizing the nature of the new particles in case of discovery by exploiting beam polarization. The SUSY EW sector, comprising sleptons, can also account for the long-standing discrepancy of $(g − 2)\mu$. Feasibility studies focusing on the relevant parameter region have been reported in the past year [46], showing good complementarity between HL-LHC and high-energy electron–positron colliders. Sensitivity to staus at lepton colliders would again be complementary to $pp$ colliders in case of compressed scenarios, with substantial dependency on the assumptions on $\tilde{\tau}$ handedness and the beam polarization conditions. A multi-TeV muon collider would push the sensitivity up to half the center-of mass energy [34].

A systematic study of the potential of lepton and hadron colliders for $R$-parity violating (RPV) SUSY scenarios has not been attempted. RPV models might lead to very diverse signatures depending on which couplings are different from 0 and on their strength. The lightest neutralino, as LSP, would decay into SM particles so that final state events present high lepton and/or jet multiplicities and modest or no missing transverse momentum. If RPV couplings are small, particles might travel macroscopic distances before decaying and be long-lived. Searches for high-mass long-lived particles (e.g., gluinos and top squarks) at high-energy $pp$ colliders can exploit the capability of reconstructing unconventional signatures such as massive displaced vertices or displaced leptons similarly to current and planned searches at the LHC and HL-LHC, respectively. Similar techniques can be adopted at lepton colliders for EWinos. To illustrate the potential to discover low-mass SUSY particles decaying non-promptly and as such eluding LHC constraints, interesting prospect studies have been made for O (100 GeV) center-of-mass energy $e^+e^-$ colliders [47] and at $ep$ colliders [48].

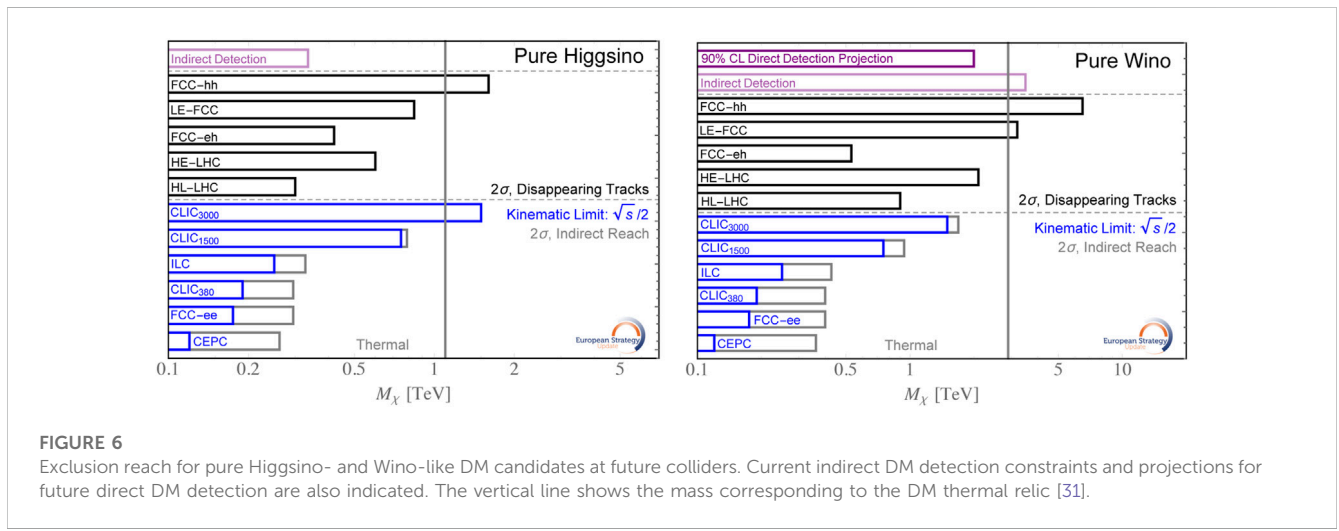

**FIGURE 6**
Exclusion reach for pure Higgsino- and Wino-like DM candidates at future colliders. Current indirect DM detection constraints and projections for future direct DM detection are also indicated. The vertical line shows the mass corresponding to the DM thermal relic [31].

# 7 Dark matter

Collider experiments could be sensitive to many possible thermal freeze-out scenarios which assume that DM was produced through interactions of unknown nature with SM particles in the early universe. If DM is part of an extended BSM sector and carries SM charges, a mechanism should exist, as in SUSY, to make it stable or very weakly interactive (WIMP). In case of DM being part of a richer hidden sector, several new massive particles might arise, and one or more could mix with SM particles. A hidden sector that contains DM is generically called a dark sector and can be connected to the SM by small but non-zero couplings through a mediator. The operator interacting or mixing with it is referred to as a portal.

Depending on the model assumptions, the nature of DM and the new-physics phenomenology could be profoundly different. For instance, dark sectors might be characterized by an abundance of feebly interacting particles. Feasibility studies on these scenarios are reported in Section 8.

WIMP DM is invisible to detectors due to the weak strength of its interaction with SM particles, and hence the main signature at colliders is the missing transverse momentum carried by the DM particle. Consequently, searches focus on the associated production of the undetectable DM with visible SM particle(s) like one (or more) jet(s), a $Z$ boson, a photon, or a Higgs boson. Additional BSM mediators can lead to a variety of even more complex collider signatures in visible channels, i.e., involving heavy-flavor quarks.

A straightforward model of DM thermal relic is that of a massive particle with EW gauge interactions only. The case of spin-1/2 particles transforming as doublets or triplets under SU(2) symmetry is considered an excellent benchmark model for future colliders. The production rate of the charged state in the DM EW multiplet is high, but it decays into the invisible DM plus a soft undetectable pion. The sensitivity to these models, usually referred to as Higgsino and Wino, respectively, is summarized in Figure 6.

The direct reaches through the so-called disappearing track analyses are compared with indirect reaches at lepton colliders, derived from the sensitivity to the EW parameters $W$ and $Y$. FCC-hh can conclusively test the hypothesis of thermal DM for both the Higgsino and Wino scenarios, while CLIC$_{3000}$ could cover in full the Higgsino case. A 3 TeV muon collider would reach masses slightly lower than CLIC$_{3000}$ for the Wino case, while a 10 TeV machine would be competitive with the FCC-hh [34]. As usual, several caveats must be considered when comparing these projections. For instance, projections for future direct DM detection might suffer from uncertainties on the Wino-nucleon cross section, whilst indirect constraints might suffer from unknown halo-modeling uncertainties. More details can be found in Ref. [31] and references therein.

If DM belongs to a richer BSM sector, the phenomenology might be very diverse. Simplified models are therefore used as benchmarks for collider searches to minimize the number of unknown parameters: a single mediator is introduced, which is either a new BSM particle or a SM particle such as the Higgs boson or the Z boson. In the models considered by the EPSSU studies, based on widely accepted benchmark proposals [49], the DM particle is a massive Dirac fermion ($\chi$), and the mediator is either a spin-1 (axial-vector) or a spin-0 (scalar) BSM particle. Figure 7 (left) reports the $2\sigma$ sensitivity on the mediator mass of collider experiments for axial-vector models. Results are strongly dependent on the choice of couplings (indicated in the figure), and hence it is difficult to compare among accelerator projects.

The sensitivity at $pp$ colliders is driven by dijet and monojet searches, which decreases if couplings to quark decrease. Lepton colliders might reach reasonably high mediator masses through mono-photon analyses, so the achievable sensitivity depends on the strength of the mediator coupling to leptons. Similar results to those of CLIC$_{3000}$ can be achieved by a MuC of the same center-of-mass energy, with sensitivity provided by mono-photon and mono-$W$-boson analyses [30].

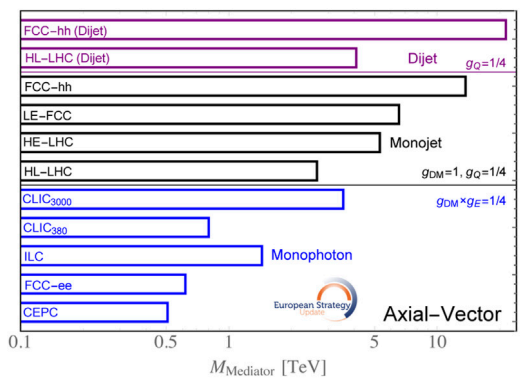 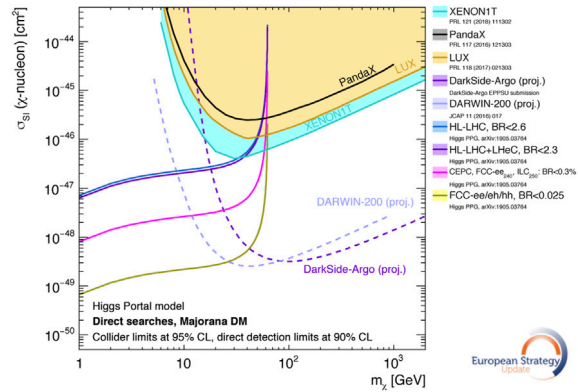

FIGURE 7
Left: Exclusion reach at 95% CL for axial-vector simplified models at future colliders assuming a DM mass of $M_{DM}$ =1 GeV. Right: Results from searches for invisible decays of the Higgs boson, assuming a Majorana-type DM, compared to constraints of current and future direct detection experiments on the spin-independent WIMP–nucleon scattering cross section [31]. The region where the irreducible background from neutrinos may hide a DM signal, usually labeled as the "neutrino floor," is not reported. For further reading, a recent paper on the subject is given in Ref. [50].

Mediators could also be SM particles, and a remarkable example is given by models where the Higgs boson acts as a mediator (or portal). If the DM mass is below half of the mass of the Higgs boson, the latter could decay into a DM pair. As such, precision measurements of the branching ratio (BR) of the Higgs boson decaying into invisible particles can be turned into exclusion limits on the spin-independent WIMP–nucleon scattering cross section. This is illustrated in Figure 7 (right): 90% CL limits for a simplified model with the Higgs boson decaying to Majorana DM particles are compared to current and future DM direct detection experiments. Low-energy $e^+e^-$ colliders are particularly competitive in this scenario, thanks to unprecedented precision expected in measuring Higgs couplings, whilst hadron colliders remain competitive thanks to the large datasets and high production rates.

## 8 Feebly interacting particles

BSM theories extending the SM with a hidden sector populated by feebly interacting particles (or FIPs) are gaining significant attention as they can provide, depending on the model's implementation, an explanation for the origin of neutrino masses, matter–antimatter asymmetry in the universe, and cosmological inflation, as well as insights into the EWK hierarchy and the strong CP problem. A comprehensive overview of the vast program at both current and future collider-based, fixed-target, and beam-dump experiments can be found in Refs [31, 51]. In this review, the focus is on the minimal portal framework introduced in the aforementioned references. In these models, the FIPs, which are not charged under the SM gauge groups, interact with the SM through portals that can be classified based on the type and dimension of the mediator. The most studied cases, listed as follows according to the operator's spin, are the vector, Higgs, axion, and neutrino portals:

where $F'_{\mu\nu}$ is the field strength for the dark vector, which mixes with the hypercharge field strength $B^{\mu\nu}$; $S$ (sometimes referred to as $\phi$) is the *dark Higgs*, a new scalar singlet that couples to the SM Higgs doublet; $a$ is the axion or axion-like particle that couples to gauge and fermion fields (pseudoscalar portal, where $f_a$ is the axion decay constant); and $N$ is a *heavy neutral lepton* (HNL) that couples to the SM left-handed leptons.

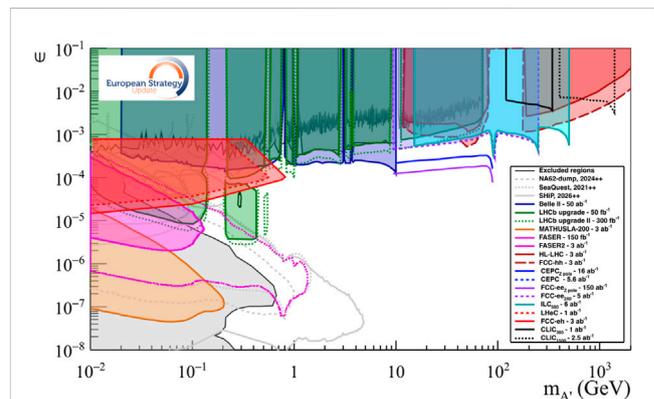

FIGURE 8
Exclusion reach for dark photons at various experiments as a function of the dark photon mass $m_{A'}$ and the mixing parameter between the dark and SM photon, $\epsilon$. Exclusion limits are computed at 95% CL in the case of FCC-ee, FCC-hh, and CepC, while the contour for the FC-eh corresponds to the observation of 10 signal events. All other limits are computed at 90% CL [31].

In the minimal vector portal, the interaction between the SM and the hidden sector takes the form of a kinetic mixing between one dark and one visible Abelian gauge boson. In selected realizations of the portal, the new U (1) gauge boson in the hidden sector is a dark photon, $A'$, either massive or massless, with $\epsilon$ being the mixing coupling parameter between the dark and ordinary photon. Figure 8 presents the sensitivity of various experiments, demonstrating that future colliders can probe the MeV to TeV mass region, compatible with the hypothesis of DM as a thermal relic. Through searches for Drell–Yan production, $pp \rightarrow A' \rightarrow \ell^+\ell^-$, high-energy hadron colliders explore scenarios with large couplings and heavy dark photons, with the FCC-hh yielding a sensitivity to $\epsilon$ as small as $10^{-3}$ for masses in the 10–100 GeV range [52] and thus complementing the reach of future LHCb upgrades [4]. An integrated program of precision measurements of the $Z$ boson properties and of direct searches exploiting the radiative return processes ($e^+e^- \to \gamma A'$) enables electron–positron colliders to significantly extend the sensitivity to lower couplings and lower masses, with FCC-ee probing couplings close to $\epsilon \sim 10^{-4}$ for dark photon masses below the $Z$ mass. For higher masses, the sensitivity of circular and linear $e^+e^-$ colliders with similar luminosities is comparable. Searches for long-lived dark photons produced in $ep \to eA'$ and decaying into two charged fermions enable the FCC-eh to probe for masses below 1 GeV and couplings in between $10^{-5}$ and $10^{-3}$ [53], filling the gap between LHCb, future $e^+e^-$ and $pp$ colliders, and low-energy experiments. Recent results from searches at a muon collider are presented in Ref. [30], while Ref. [54] offers a comprehensive review on more general dark-photon models and corresponding searches. It is noted that hadron and lepton colliders could offer significant sensitivity to non-minimal models where dark photons are produced through BSM Higgs decays, as shown for the HL-LHC in Ref. [4]. A detailed discussion of the sensitivity to non-minimal scenarios is, however, outside the scope of this review.

| Portal | Coupling |
|---|---|
| Vector (dark vector, $A_\mu$) | $-\frac{\epsilon}{2\cos\theta_W} F'_{\mu\nu} B^{\mu\nu}$ |
| Scalar (dark Higgs, $S$) | $(\mu S + \lambda_{HS} S^2) H^\dagger H$ |
| Pseudo-scalar (axion, $a$) | $\frac{a}{f_a} F_{\mu\nu} \tilde{F}^{\mu\nu}, \frac{a}{f_a} G_{i,\mu\nu} \tilde{G}_i^{\mu\nu}, \frac{\partial_\mu a}{f_a} \bar\psi \gamma^\mu \gamma^5 \psi$ |
| Fermion (sterile neutrino, $N$) | $y_N LHN$ |

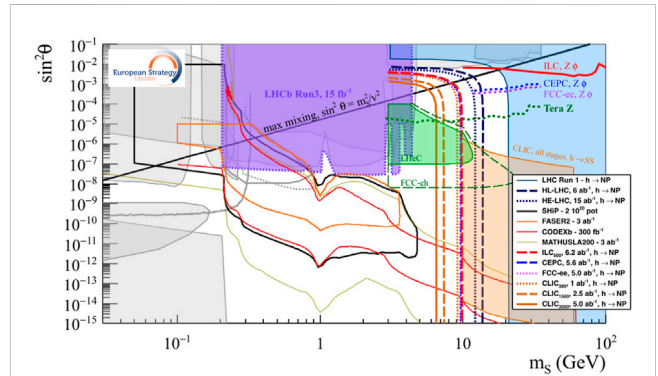

FIGURE 9
Exclusion reach in the $m_S - \sin^2\theta$ plane at various experiments. The symbols $S$ and $\phi$ are used interchangeably to identify the singlet. Exclusion limits are computed at 95% CL in the case of FCC-ee, FCC-hh, ILC, CLIC, and CepC, while all other limits are computed at 90% CL [31].

The minimal dark Higgs model originates from the extension of the SM Higgs sector by the addition of a scalar singlet which mediates the interaction between the SM Higgs boson and the dark sector. In the context of general extended Higgs models, the Lagrangian contains a term proportional to $\sin\theta$ (referred to as $\sin\gamma$ in Section 5), enabling the mixing between the SM Higgs and the new particle $S$ associated with the singlet field, with mixing angles $\theta$. The Lagrangian also contains a term proportional to $\lambda_{HS} S^2$, leading to the coupling between the $h$ and two $S$ particles. If either of the couplings $\sin\theta$ or $\lambda_{HS}$ is non-zero, a rich phenomenology is expected. If the new scalar mixes with the SM Higgs boson, $S$ can either be produced like a SM Higgs boson or originate from exotic decays of the SM Higgs boson. The corresponding cross sections and branching fractions would depend on the mixing angle. Once produced, the new scalar could decay like a SM Higgs boson, with probabilities reduced by $\sin\theta$, and into the SM Higgs boson itself if $m_S > 2m_h$. In the no-mixing scenarios, $S$ can only be pair-produced through an off-shell or on-shell Higgs boson. The new scalar is stable in the no-mixing minimal models, leading to signatures with missing transverse momentum. The region of parameter space with larger couplings, $\sin^2\theta \geq 10^{-4}$, is explored by searches for the associated production of $ZS$ conducted using the recoil technique at $e^+e^-$ colliders. The Tera-Z configuration of the FCC-ee extends the reach in couplings by one order of magnitude for masses in between a few GeV and half of the $Z$ boson mass by exploring the exotic decays $Z \to \ell^+\ell^- S$. Precision measurements of the Higgs couplings place constraints on the mass over a large range of $\sin^2\theta$ values: for a fixed luminosity, $e^+e^-$ colliders yield a better sensitivity than those proton–proton machines included in this study, with CLIC$_{3000}$ covering masses as low as 6 GeV for $\sin^2\theta \leq 10^{-5}$. Searches for $h \to SS$ in visible final states at the FCC-eh allow the experiments to probe intermediate values of $\sin^2\theta$ for masses $m_S$ between 3 and 30 GeV, while similar analyses at CLIC extend the sensitivity to lower values of the couplings for masses between 10 and 60 GeV. These searches assume $\lambda_{HS} \sim 10^{-3}$, corresponding to the level of precision on the measurements on the SM Higgs coupling expected at future $e^+e^-$ colliders. A summary is presented in Figure 9, which also includes the relation between the relevant parameters under the maximal mixing assumption in relaxion models as they exhibit a similar phenomenology via relaxion-Higgs mixing.

Axion-like particles (ALPs, $a$) are gauge-singlet pseudoscalar particles with derivative couplings to the SM. ALPs can mediate the interactions between the SM and the hidden sector by coupling to photons, gluons, $W$ and $Z$ bosons, and fermions. The interactions with the Higgs boson are suppressed since there is no dimension-5 operator at the tree level in the models considered here. At high-energy colliders, ALPs emerge from either resonant production or from exotic decays of the $Z$ or Higgs bosons ($Z \to a\gamma, h \to aZ, aa$). In addition, they can be produced via vector–boson fusion at $pp$ colliders and in association with a gauge or Higgs boson at lepton colliders ($e^+e^- \to aX$ with $X = \gamma, Z, h$). In $ep$ machines, the incoming electron interacts with a photon from the proton, leading to $e^-\gamma \to e^-a$. For ALP masses, $m_a$, below the $Z$ mass, the dominant decay modes are into gluons and photons, where the latter has received the most attention to date. Results from recent searches are therefore presented as a function of the ALP mass and coupling to photons (Figure 10). Thanks to excellent sensitivity to the process $e^+e^- \to Z \to a\gamma(\gamma\gamma)$, the Tera-Z configuration of the FCC-ee reaches the best sensitivity for ALP masses between the ~1 GeV and the $Z$ mass, probing couplings $g_{a\gamma\gamma}$ as small as $10^{-8}$. Searches for the same rare decay at the FCC-hh probe have a similar mass range but with somewhat limited coverage in couplings, as expected. On the other hand, hadron colliders offer excellent sensitivity to scenarios where, e.g., the ALP originates from rare Higgs decays [55]. The $e^+e^-$ linear colliders extend the reach at larger masses thanks to their higher center-of-mass energy and probe couplings as small as $10^{-5}$. Experiments at $ep$ colliders have a reach similar to those of low-energy $e^+e^-$ linear colliders by searching for the existence of $e\gamma \to ea$. A detailed overview of the subject, comparing the reach at various machines, is provided in Ref. [55], while Ref. [56] goes into the details of searches at the FCC-ee, exploring all combinations of ALP production modes with visible and invisible decay modes, including those associated with long-lived ALPs. Recent studies at $\sqrt{s} = 10$ TeV muon collider, performed using a modified nomenclature, indicate a discovery potential up to an effective energy scale $\Lambda$ of 238 TeV, where $\Lambda$ controls the strength of the interactions [30].

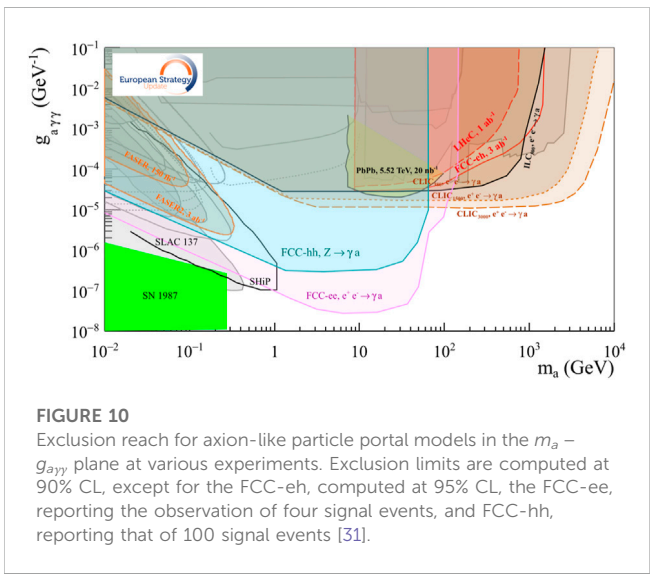

**FIGURE 10**
Exclusion reach for axion-like particle portal models in the $m_a$ – $g_{a\gamma\gamma}$ plane at various experiments. Exclusion limits are computed at 90% CL, except for the FCC-eh, computed at 95% CL, the FCC-ee, reporting the observation of four signal events, and FCC-hh, reporting that of 100 signal events [31].

Heavy neutral leptons (also referred to as heavy neutrinos or sterile neutrinos) are one of the most promising extensions of the SM to generate the light neutrino masses observed in neutrino oscillation experiments. At colliders, HNL can emerge from leptonic decays of the $W$, $Z$, and Higgs bosons with a probability proportional to the mixing with the SM neutrinos, where the mixing angles and their magnitude can be expressed as $\theta_i = \frac{y^*_{\nu_i}}{\sqrt{2}} \frac{VEV}{M}$ and $|\Theta^2| = \sum_i \theta^2$ ($y^*_{\nu_i}$ are the Yukawa couplings, while $M$ is the sterile neutrino degenerate mass, and $VEV$ is the vacuum expectation value). Additional production mechanisms include the $W$-exchange $t$-channel at both $e^+e^-$ and $ep$ colliders ($eq \to Nq$; $e^+e^- \to N\nu$) and $\gamma W$ fusion at $e^+e^-$ machines ($e\gamma \to NW$). Subsequent decays of $N$ occur via emission of a $W$, $Z$, $h$ boson, if kinematically allowed. Depending on the value of the couplings and masses, the decay may be prompted or delayed. This rich phenomenology [57] offers opportunities for both direct searches for these new states, e.g., $Z \to N\nu$ or $W \to N\ell$, as well as indirect searches through precision measurements of the gauge and Higgs bosons' properties. The $h \to WW$ channel, e.g., is used to place indirect constraints on $h \to N\nu$. Figure 11 (left) shows that the best sensitivity to the mixing parameter between the electron neutrino and HNL in the region between a few GeVs and the $Z$ mass is yielded by a combination of the conventional and of the displaced-vertex searches performed at the FCC-ee at the $Z$-pole. For larger masses, FCC-eh provides the best sensitivity for couplings as small as $10^{-6}$ through searches for lepton-flavor-violating decays.

Further studies in Ref. [57] are presented in Figure 11 (right), where $\theta_e = \theta_\mu = \theta_i$, $\theta_i \neq 0$, and $\theta_\tau = 0$. In these models, the sensitivity of indirect searches pushes the sensitivity of the FCC-ee to the TeV scale. Muon colliders could complement the FCC-ee capability in hierarchical scenarios where the mixing to the second generation is dominant. Finally, the recent work in Ref. [56] provides in-depth considerations about the reach of searches for long-lived HNL, the potential to discriminate between the Dirac and Majorana hypotheses, to measure the mass, and to probe regions of parameter space consistent with leptogenesis. The experimental sensitivity to heavy neutrinos embedded in UV complete theories, like supersymmetry or type III 2HDM, is discussed, e.g., for the ILC, in Ref. [27].

# 9 Rare processes and indirect BSM physics searches

The presence of new phenomena at a high-energy scale can impact the production rate of processes that are otherwise very rare in the SM. Examples already mentioned in this review are the anomalies in measurements such as $R_K$ and $R_K^*$ at LHCb [15] that can be explained by the presence of LQ or $Z'$. Flavor-changing neutral current effects in the heavy-quarks and gauge boson sectors are another case extensively investigated at future colliders. Prospect studies can be found in (32) and references therein, with the expected sensitivity of future lepton collider proposals at critical production thresholds complementary to that of hadron colliders at very high luminosities.

In the absence of evidence for new physics, the formalism of EFT can be adopted as a global framework to perform model-independent searches. Two effective field theory approaches are considered here. The first one, the Standard Model EFT or SMEFT, extends the SM with operators ($\mathcal{O}_i$) at higher canonical dimension "$d$," constructed as combinations of SM fields, invariant under the Lorentz and SM gauge symmetries. If lepton and baryon number conservation is imposed to reduce the otherwise very large number of possible new operators, the first corrections to the SM are provided by operators of dimension six. BSM physics at energies below $\Lambda$ is then described by a Lagrangian $\mathcal{L} = \mathcal{L}_{SM} + \mathcal{L}_{BSM}$, where $\mathcal{L}_{BSM} = \sum_{d>4} \frac{1}{\Lambda^{d-4}} \mathcal{L}_d$ and $\mathcal{L}_d = \sum_i c_i^{(d)} O_i^{(d)}$. The Wilson coefficients $c_i^{(d)}$ depend on the structure of new physics. Furthermore, the first corrections to the SM are provided by operators of dimension six if lepton and baryon number conservation is imposed. Since BSM-induced corrections to the SM parameters can be grouped into sets of models, any deviations of the SM parameters from their expectations could provide an indication about $\frac{c_i}{\Lambda^2}$. Thanks to its linearized Lagrangian, SMEFT is an excellent tool to probe for weakly coupled theories. Recent studies based on global fits to SMEFT operators are documented in Refs [30, 38] and shown in Figure 12: these selected results, shown as a relative improvement compared to the HL-LHC results, indicate that BSM scales between 1 and several tens of TeVs can be probed at future colliders under the assumption of $c_i \sim 1$. Precision measurements at future lepton colliders, in particular where $Z$-pole runs are planned, contribute substantially to the extraction of EW but also triple-gauge coupling parameters. Diboson and $Zh$, $h \to b\bar{b}$ measurements in the boosted regime are among the probes most relevant for high-energy $pp$ colliders. For muon colliders, high-energy measurements in two-to-two fermion processes as well as single-Higgs and di-Higgs precision measurements have been considered. Even though the SMEFT provides a consistent framework to describe the impact of BSM physics, it is important to highlight that the results of the global fit depend on the choice of operators, basis, selected observables and their correlations, experimental and theoretical systematic uncertainties, etc. A detailed discussion on the subject can be found in Ref. [58]. For instance, off-diagonal flavor structures are not considered. If the imposed benchmark flavor symmetry is relaxed, top-quark FCNC interactions can be incorporated [59] into the SMEFT framework.

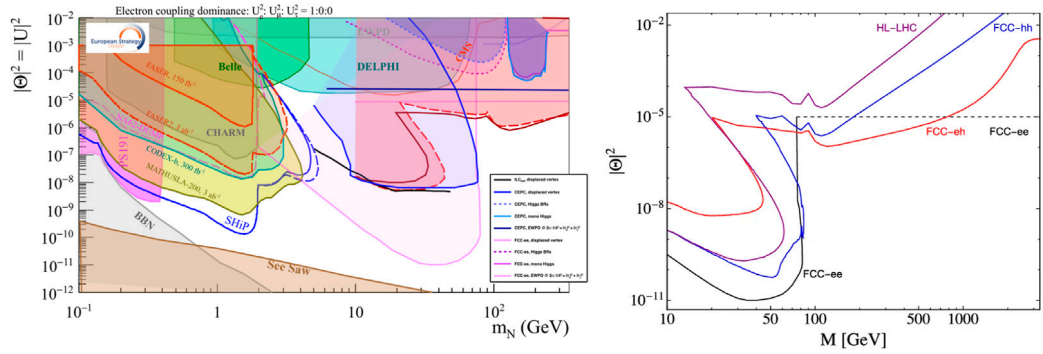

FIGURE 11
Left: Exclusion reach for minimal heavy lepton portal models assuming electron-coupling dominance in the $|\Theta|^2$– $m_N$ plane at various experiments. Exclusion limits are computed at 90% CL [31]. Right: Comparison of exclusion reaches at 90% CL from searches at the HL-LHC, FCC-hh, and FCC-eh and precision measurements at the FCC-ee [57].

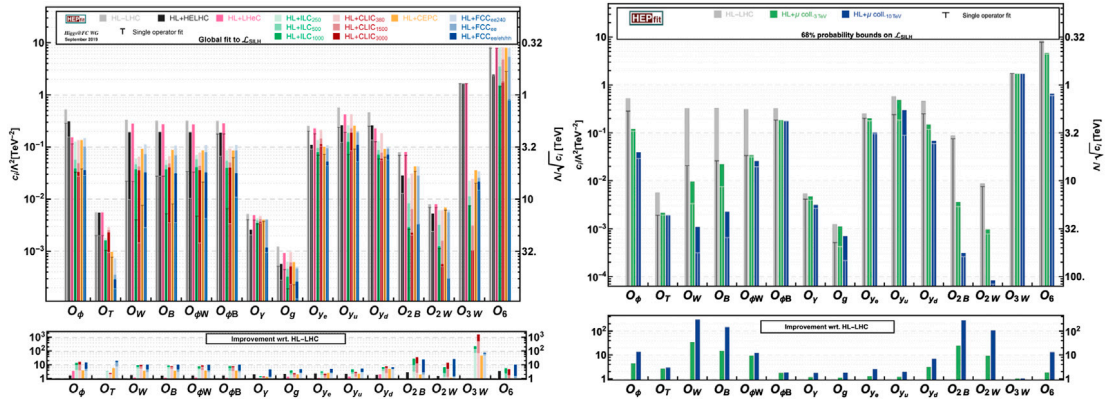

FIGURE 12
Probability reach for the Wilson coefficients computed at 68% CL from the global fit (solid bars). The vertical "T"-shaped lines report the results obtained if only one operator is generated by the UV dynamics. Left: Reach of all options considered in Ref. [38]. Right: Sensitivity of the muon collider at both $\sqrt{s}$ = 3 and 10 TeV, compared to that of the HL-LHC [30].

The second EFT approach is the so-called Higgs EFT (HEFT). In this framework, the Higgs boson is not required to belong to an exact $SU(2)_L$ doublet, and the physical Higgs and the three EW Goldstone bosons are treated as independent objects. The physical Higgs is typically assigned to a singlet representation of the SM gauge groups.

The HEFT, with a non-linear realization of the EWSB, offers the most general description of the Higgs couplings, and it is suitable to investigate a large set of distinct theories, including composite Higgs, and scenarios with modified trilinear Higgs couplings. Even though HEFT is outside the scope of this review, the reader is invited to find detailed comparative overviews of SMEFT and HEFT in Refs [60, 61].

## 10 State of the art of the proposed colliders

The broad and ambitious science program presented in the previous sections, and references therein, depends critically upon the performance of the accelerators and experiments, including both the instrumentation and software and computing elements. Advances in theoretical methods are also essential for the full exploitation of these machines' discovery potential. The technology needed to realize these projects might not exist yet, and cutting-edge and vigorous R&D is therefore being pursued by the global HEP experimental community. Extensive studies on detector concepts are ongoing within the broader worldwide detector R&D programs, as recently presented in Ref. [62]. A succinct summary of the status of the machines considered in this paper, as well as recently proposed modifications and upgrades, is presented in the following paragraphs. The reader is invited to consult dedicated reviews to learn about the latest developments in advanced accelerator techniques, instrumentation, software and computing, and theoretical calculations and methods.

The technical design report (TDR) of the ILC, released in 2013 [64], focused on the 250–500 GeV option (with a possible energy upgrade to 1 TeV). The recent comprehensive report submitted to Snowmass [27] reviews in detail the accelerator design, proposes new luminosity and energy running conditions (including technology options for multi-TeV upgrades), and presents robust solutions to deliver electron and positron beams in the energy region of the Higgs boson. In addition, it updates the proposal made in the detector TDR [63] for two detectors at the interaction region, the SiD and the ILD detector, and outlines that further R&D is needed. These new detector designs have been carried out at the level of a conceptual design report (CDR). In 2020, ICFA approved the formation of the ILC International Development Team as the first step toward the preparatory phase of the ILC project, with a mandate to make preparations for the ILC Pre-Lab in Japan [64] while waiting for a decision by Japan about hosting the facility. If approved, the machine is expected to deliver physics data in the mid-2030s.

As a direct response to a high-priority request from the 2013 update of the European Strategy for Particle Physics, CERN developed the design for the Future Circular Collider. The CDR for the FCC housed in a 100-km-long tunnel at CERN was delivered in 2018 with preliminary cost estimates and feasibility assessments [6, 9, 42]. Updates were presented within the 2018 process for the European Strategy and, more recently, in the context of the Snowmass community planning process. The latest proposals [65–67] include a phased approach with an $e^+e^-$ machine at various center-of-mass energies (including at the Higgs mass), followed by the 100 TeV $pp$, the $ep$, and the heavy ion programs. Under the auspices of CERN, the FCC Collaboration is also considering a tunnel of 91 km. Assuming a timely completion of the R&D for the FCC-ee, start of operations is expected in the 2040s, with data taking lasting till 2060. The FCC-hh is scheduled to run between 2070 and 2090. The program may be modified to focus on the $pp$, the $ep$, and the heavy ion programs if an $e^+e^-$ collider other than the FCC-ee is approved for construction. Two to four experiments could be hosted by the FCC complex at each stage.

After the release of the CepC CDR and subsequent studies documented in Ref. [68], the CepC accelerator study group entered the TDR phase, expected to be completed by the end of 2022. Meanwhile, an update to the design of the CepC and SppC, including a target center-of-mass energy of 125 TeV for $pp$ collisions, is discussed in Ref. [69] and Ref. [70], respectively. According to the currently envisioned schedules, the CEPC (SppC) starts operations in the mid-2030s (2050s).

The CLIC CDR [7], documenting the 3 TeV machine, dates back to 2012, while a project implementation plan, PIP [71], was finalized in 2018 and included the option for the 380 GeV running. The recent Snowmass report [72], building on both the CDR and the PIP, describes recent achievements in accelerator design, technology development, system tests, and beam tests for CLIC, demonstrating that the performance goals are realistic. In addition, results from ongoing R&D are likely to allow for further upgrades, both in $\sqrt{s}$ and instantaneous luminosity. Assuming project approval in 2028 (after the next ESPP), construction can start in ~2030 and operations ~7 years later.

The option of a muon collider has gained substantial interest in the past two years. Documented in Ref. [73] is the latest proposal for a muon collider with three tentative target center-of-mass energies: 3, 10, and 14 TeV. Other energy conditions are also explored, e.g., operations at $\sqrt{s}$ =125 GeV. It is recognized that the muon collider is not as mature as the other high-energy lepton collider options listed previously. However, no major technical limitations are identified to date, and the outlined R&D path to address the remaining challenges makes the 3 TeV viable, with start of data taking in the mid-2040s.

As mentioned in Section 2, the newly proposed $C^3$ linear accelerator [17] benefits from recent advances that increase the efficiency and operating gradient of a normal conducting accelerator and provides a solution to a multi-TeV machine. The current proposal is for a compact 8-km-long cold copper-distributed coupling complex that could fit on the Fermilab site. The Snowmass 2021-contributed paper [18] instead discusses the Higgs-Energy LEptoN (HELEN) $e^+e^-$ linear collider, based on advances in superconducting radio frequency technology, but with potential cost and AC power savings and smaller footprint (relative to the ILC). If the ILC cannot be realized in Japan in a timely fashion, the HELEN collider would be another viable option for a Higgs factory in the U.S.

## 11 Conclusion

Several collider projects have been proposed and discussed in recent years. Each proposal offers compelling opportunities for precision measurements and searches for new physics, albeit carrying challenges in accelerator, detector, and computing technologies. The ESPPU and the Snowmass/P5 processes have outlined future prospects at linear and circular $e^+e^-$, high-energy $pp$, $ep$, and high-energy $\mu^+\mu^-$ colliders, along with their capability to solve long-standing problems, such as the understanding the EWSB mechanism, the origin and nature of dark matter, the flavor problem, the origin of neutrino masses, the strong CP problem, and baryogenesis. This review has briefly summarized the outcomes of those processes, resulting from the huge combined effort of the theory and experimental particle physics communities during the last 5 years, in the context of direct and indirect searches for new physics. Well-motivated BSM scenarios have been considered to provide quantitative comparisons between the reach of different proposed projects. The emerging picture shows that, while there are excellent chances for fundamental discoveries at the HL-LHC, the datasets might not be sufficient to fully characterize new physics if deviations from the SM are found, and future colliders will make this possible. Each future collider offers exciting prospects to enable the exploration of the unknown beyond the HL-LHC, and the realization of one or more of the proposed accelerator projects in the next decades should be strongly pursued by the scientific community to guarantee unique advancements in the understanding of the laws of nature.

## Author contributions

AC and MD reviewed the existing published material on searches for new physics at future colliders and summarized it in this review paper. All authors contributed to the article and approved the submitted version.

## Funding

MD was partially funded by STFC (PP(E)CG 2022-2025, ST/W000466/1), United Kingdom, and by the CHIST-ERA grant, CHIST-ERA-19-XAI-009, under Horizon 2020. AC was supported by Fermi Research Alliance, LLC, under Contract No. DE-AC02- 07CH11359, with the U.S. Department of Energy, Office of Science, Office of High-Energy Physics.

## Acknowledgments

The authors wish to thank the thousands of scientists that in the past years have worked strenuously on the proposals considered in this report for their dedication and vision. The authors thank Carl Gwilliam for useful discussions and the thorough review of this manuscript and Oliver Fischer for useful discussions about future colliders' reach for heavy neutral leptons.

## Conflict of interest

The authors declare that the research was conducted in the absence of any commercial or financial relationships that could be construed as a potential conflict of interest.

## Publisher's note

All claims expressed in this article are solely those of the authors and do not necessarily represent those of their affiliated organizations, or those of the publisher, the editors, and the reviewers. Any product that may be evaluated in this article, or claim that may be made by its manufacturer, is not guaranteed or endorsed by the publisher.

## References

1. Cepeda M, Gori S, Ilten P, Kado M, Riva F, Abdul Khalek R, et al. Report from working group 2 Higgs physics at the HL-LHC and HE-LHC. *CERN Yellow Rep Monogr* (2019) 7:221. arXiv 1902.00134. doi:10.23731/CYRM-2019-007.221

2. Azzi P, Farry S, Nason P, Tricoli A, Zeppenfeld D, Abdul Khalek R, et al. Report from working group 1 standard model physics at the HL-LHC and HE-LHC. *CERN Yellow Rep Monogr* (2019) 7:1. arXiv 1902.04070. doi:10.23731/CYRM-2019-007.1

3. Agostini P, Aksakal H, Alekhin S, Allport PP, Andari N, Andre K, et al. The large Hadron\textendash{}Electron collider at the HL-LHC. *J Phys G* (2021) 48:110501. arXiv 2007.14491.

4. Vidal X, D'Onofrio M, Fox P, Torre R, Ulmer KA, Aboubrahim A, et al. Report from working group 3 beyond the standard model physics at the HL-LHC and HE-LHC. *CERN Yellow Rep Monogr* (2019) 7:585. arXiv 1812.07831.

5. Cerri A, Gligorov VV, Malvezzi S, Martin Camalich J, Zupan J, Akar S, et al. Report from working group 4 opportunities in flavour physics at the HL-LHC and HE-LHC. *CERN Yellow Rep Monogr* (2019) 7:867. arXiv 1812.07638. doi:10.23731/CYRM-2019-007.867

6. Abada A, Abbrescia M, AbdusSalam SS, Abdyukhanov I, Abelleira Fernandez J, Abramov A, et al. HE-LHC: The high-energy large hadron collider. *Eur Phys J ST* (2019) 228:1109–382. doi:10.1140/epjst/e2019-900088-6

7. Aicheler M, Burrows P, Draper M, Garvey T, Lebrun P, Peach K, et al. A multi-TeV linear collider based on CLIC technology CLIC conceptual design report. *CERN Yellow Rep Monogr* (2012) 7. doi:10.5170/CERN-2012-007

8. CEPC Study Group, (2018), CEPC conceptual design report: Volume 1 - accelerator, arXiv 1809.00285.

9. Abada A, Abbrescia M, AbdusSalam SS, Abdyukhanov I, Abelleira Fernandez J, Abramov A, et al. FCC-hh: The hadron collider future circular collider conceptual design report volume 3. *Eur Phys J ST* (2019) 228:755–1107. doi:10.1140/epjst/e2019-900087-0

10. Arduini G, Brüning O, Klein M Energy frontier DIS at CERN: The LHeC and the FCC-eh, PERLE. *Pos DIS* (2018). 183:8. doi:10.22323/1.316.0183

11. Willeke F *BNL-221006-2021-FORE* (2021).

12. Abada A, Abbrescia M, AbdusSalam SS, Abdyukhanov I, Abelleira Fernandez J, Abramov A, et al. HE-LHC: The high-energy large hadron collider: Future circular collider conceptual design report volume 4. *Eur Phys J ST* (2019) 228:1109–382. doi:10.1140/epjst/e2019-900088-6

13. Long K, Lucchesi D, Palmer M, Pastrone N, Schulte D, Shiltsev V. Muon colliders: Opening new horizons for particle physics. *Nat Phys* (2021) 17:289–92. arXiv 2007.15684. doi:10.1038/s41567-020-01130-x

14. Muon G, Abi B, Al-Kilani S, Allspach D, Alonzi L, Anastasi A, et al. Measurement of the positive muon anomalous magnetic moment to 0.46 ppm. *Phys Rev Lett* (2021) 126:141801. arXiv 2104.03281. doi:10.1103/PhysRevLett.126.141801

15. Aaij R, Abellán Beteta C, Ackernley T, Adeva B, Adinolfi M, Afsharnia H, et al. Test of lepton universality in beauty-quark decays, *Nat Phys*. 18, 277 (2022), arXiv 2103.11769. doi:10.1038/s41567-021-01478-8

16. Bhat P, Jindariani S, Ambrosio G, Apollinari G, Belomestnykh S, Bross A, et al. (2022), Future collider options for the US, arXiv 2203.08088.

17. Bai M, Barklow T, Bartoldus R, Breidenbach M, Grenier P, Huang Z, et al. (2021), C$3$: A "Cool" route to the Higgs boson and beyond, arXiv 2110.15800.

18. Belomestnykh S, Bhat PC, Grassellino A, Checchin M, Denisov D, R. L. Geng, et al. (2022), Higgs-Energy LEptoN (HELEN) Collider based on advanced superconducting radio frequency technology, arXiv 2203.08211.

19. Ahmad M, Alves D, An H, An Q, Arhrib A. *CEPC-SPPC preliminary conceptual design report. 1. Physics and detector*. IHEP-CEPC-DR-2015-01 (2015).

20. Ariga A, Ariga T, Boyd J, Cadoux F, Casper DW, Favre Y, et al. (2019), Faser: ForwArd search ExpeRiment at the LHC, arXiv 1901.04468.

21. Ahdida C, Akmete A, Albanese R, Alexandrov A, Andreini M, Anokhina A, et al. (2020), SND@LHC, arXiv 2002.08722.

22. Lubatti H, Alpigiani C, Arteaga-Velázquez JC, Ball A, Barak L, Beacham J, et al. Explore the lifetime frontier with MATHUSLA. *JINST* (2020) 15:C06026. arXiv 1901.04040. doi:10.1088/1748-0221/15/06/c06026

23. Aielli G, Ben-Haim E, Cardarelli R, Charles MJ, Cid Vidal X, Coco V, et al. Expression of interest for the CODEX-b detector. *Eur Phys J C* (2020) 80:1177. arXiv 1911.00481. doi:10.1140/epjc/s10052-020-08711-3

24. Milli Q, Yoo Y. The milliQan experiment: Search for milli-charged particles at the LHC. *PoS* **ICHEP2018** (2019) 520. arXiv 1810.06733.

25. Abelleira-Fernandez J, Akay AN, Aksakal H, Albacete JL, Alekhin S, et al. A large hadron electron collider at CERN: Report on the physics and design concepts for machine and detector. *J Phys G* (2012) 39:075001. arXiv 1206.2913. doi:10.1088/0954-3899/39/7/075001

26. Anchordoqui L, Ariga A, Ariga T, Bai W, Balazs K, Batell B, et al. (2021), The forward physics facility: Sites, experiments, and physics potential, arXiv 2109.10905.

27. Aryshev A, Behnke T, Berggren M, Brau J, Craig N, Freitas A, et al. (2022), The international linear collider: Report to Snowmass 2021, arXiv 2203.07622.

28. Langacker P The physics of heavy Z′ gauge bosons. *Rev Mod Phys* (2009) 81:1199–228. arXiv 0801.1345. doi:10.1103/RevModPhys.81.1199

29. Helsens C, Jamin D, Mangano M, Rizzo T, Selvaggi M. Heavy resonances at energy-frontier hadron colliders. *Eur Phys J C* (2019) 79:569. arXiv 1902.11217. doi:10.1140/epjc/s10052-019-7062-3

30. de Blas J, Buttazzo D, Capdevilla R, Curtin D, Franceschini R, Maltoni F, et al. (2022), The physics case of a 3 TeV muon collider stage, arXiv 2203.07261.

31. Ellis R, Beate H, de Blas J, Maria C, Christophe G, Fabio M, et al. (2019), Physics briefing book input for the European Strategy for particle physics update 2020 arXiv 1910.11775.

32. Huang G, Jana S, Queiroz F, Rodejohann W. Probing the $R_{K^{(*)}}$ anomaly at a muon collider. Phys Rev D (2022) 105:015013. doi:10.1103/PhysRevD.105.015013

33. Giudice G, Grojean C, Pomarol A, Rattazzi R The strongly-interacting light Higgs. JHEP (2007) 06:045. 0703164. doi:10.1088/1126-6708/2007/06/045

34. Aimè C, Apyan A, Attia Mahmoud M, Bartosik N, Bertolin A, Bonesini M, et al. (2022), Muon collider physics summary, arXiv 2203.07256.

35. Barducci D, Mimasu K, No J, Vernieri C, Zurita J, Enlarging the scope of resonant di-Higgs searches: Hunting for Higgs-to-Higgs cascades in 4b final states at the LHC and future colliders, JHEP 02, 2(2020), arXiv 1910.08574. doi:10.1007/JHEP02(2020)002

36. No J, Spannowsky M Signs of heavy Higgs bosons at CLIC: An $e^+e^-$ road to the electroweak phase transition. Eur Phys J C (2019) 79:467. arXiv 1807.04284. doi:10.1140/epjc/s10052-019-6955-5

37. Buttazzo D, Franceschini R, Wulzer A. Two paths towards precision at a very high energy lepton collider. JHEP (2021) 05:219. arXiv 2012.11555. doi:10.1007/JHEP05(2021)219

38. de Blas J, Cepeda M, D'Hondt J, Ellis R, Grojean C, Heinemann B, et al. Higgs boson studies at future particle colliders. JHEP (2020) 01:139. arXiv 1905.03764. doi:10.1007/JHEP01(2020)139

39. Athron P, Balazs C, Gonzalo TE, Jacob D, Mahmoudi F, Sierra C. Likelihood analysis of the flavour anomalies and **g−2** in the general two Higgs doublet model. JHEP (2022) 01:037. arXiv 2111.10464. doi:10.1007/jhep01(2022)037

40. Baer H, Berggren M, Fujii K, List J, Lehtinen SL, Tanabe T, et al. ILC as a natural SUSY discovery machine and precision microscope: From light Higgsinos to tests of unification. Phys Rev D (2020) 101:095026. arXiv 1912.06643. doi:10.1103/PhysRevD.101.095026

41. de Blas J, Franceschini R, Riva F, Roloff P, Schnoor U, Spannowsky M, et al. The CLIC potential for new physics, 3/2018 (2018), arXiv 1812.02093. doi:10.23731/CYRM-2018-003

42. Abada A, Abbrescia M, AbdusSalam SS, Abdyukhanov I, Fernandez JA, Abramov A, et al. FCC physics opportunities future circular collider conceptual design report volume 1. Eur Phys J C (2019) 79:474. doi:10.1140/epjc/s10052-019-6904-3

43. Bambade P, Barklow T, Behnke T, Berggren M, Brau J, Burrows P, et al. (2019), The international linear collider: A global project arXiv 1903.01629.

44. Battaglia M, Blaising JJ, Marshall JS, Poss S, Sailer A, Thomson M, et al. Physics performance for scalar electron, scalar muon and scalar neutrino searches at $\sqrt{s}$ = 3 TeV and 1.4 TeV at CLIC. JHEP (2013) 09:001. arXiv 1304.2825. doi:10.1007/JHEP09(2013)001

45. Baum S, Sandick P, Stengel P. Hunting for scalar lepton partners at future electron colliders. Phys Rev D (2020) 102:015026. arXiv 2004.02834. doi:10.1103/PhysRevD.102.015026

46. Chakraborti M, Heinemeyer S, Saha I. The new "MUON G-2" result and supersymmetry. Eur Phys J C (2021) 81:1114. arXiv 2104.03287. doi:10.1140/epjc/s10052-021-09900-4

47. Wang Z-S, Wang K. Long-lived light neutralinos at future Z factories. Phys Rev D (2020) 101:115018. doi:10.1103/PhysRevD.101.115018

48. Curtin D, Deshpande K, Fischer O, Zurita J. Closing the light gluino gap with electron-proton colliders. Phys Rev D (2019) 99:055011. arXiv 1812.01568. doi:10.1103/physrevd.99.055011

49. Abercrombie D, Akchurin N, Akilli E, Maestre JA, Allen B, Gonzalez BA, et al. Dark matter benchmark models for early LHC run-2 searches: Report of the ATLAS/CMS dark matter forum. Phys Dark Univ (2020) 27:100371. arXiv 1507.00966. doi:10.1016/j.dark.2019.100371

50. O'Hare C. New definition of the neutrino floor for direct dark matter searches. Phys Rev Lett (2021) 127:251802. doi:10.1103/PhysRevLett.127.251802

51. Agrawal P, Bauer M, Beacham J, Berlin A, Boyarsky A, Cebrian S, et al. Feebly-interacting particles: FIPs 2020 workshop report. Eur Phys J C (2021) 81:1015. arXiv 2102.12143. doi:10.1140/epjc/s10052-021-09703-7

52. Curtin D, Essig R, Gori S, Shelton J. Illuminating dark photons with high-energy colliders. JHEP (2015) 02:157. arXiv 1412.0018. doi:10.1007/JHEP02(2015)157

53. D'Onofrio M, Fisher O, Wang Z. Searching for dark photons at the LHeC and FCC-he. Phys Rev D (2020) 101:015020. arXiv 1909.02312. doi:10.1103/PhysRevD.101.015020

54. Fabbrichesi M, Gabrielli E, Lanfranchi G, (2020), The dark photon arXiv 2005.01515. doi:10.1007/978-3-030-62519-1

55. Bauer M, Heiles M, Neubert M, Thamm A. Axion-like particles at future colliders. Eur Phys J C (2019) 79:74. arXiv 1808.10323. doi:10.1140/epjc/s10052-019-6587-9

56. Alimena J, Bauer M, Azzi P, Ruiz R, Neubert M, Rizzi C, et al. Searches for long-lived particles at the future FCC-ee. Front Phys (2022) 10:967881. arXiv 2203.05502.

57. Antusch A, Cazzato E, Fischer O. Sterile neutrino searches at future e−e+, pp and e−p colliders. Int J Mod Phys A (2017) 32:1750078. arXiv 1612.02728. doi:10.1142/S0217751X17500786

58. de Blas J, Durieux G, Grojean C, Gu J, Paul A On the future of Higgs, electroweak and diboson measurements at lepton colliders. JHEP (2019) 12:117. arXiv 1907.04311. doi:10.1007/JHEP12(2019)117

59. Aguilar-Saavedra J, Degrande C, Durieux G, Maltoni F, Vryonidou E, Zhang C, et al. (2018), Interpreting top-quark LHC measurements in the standard-model effective field theory arXiv 1802.07237.

60. Cohen T, Craig N, Lu X, Sutherland D. Is SMEFT enough? JHEP (2021) 03:237. arXiv 2008.08597. doi:10.1007/JHEP03(2021)237

61. Brivio I, Gonzalez-Fraile J, Gonzalez-Garcia M, Merlo L. The complete HEFT Lagrangian after the LHC Run I. Eur Phys J C (2016) 76:416. arXiv 1604.06801. doi:10.1140/epjc/s10052-016-4211-9

62. ECFA Detector R&D Roadmap Process Group. ECFA detector R&D roadmap process group (2020).

63. Behnke T, Brau J. E., Foster J, Fuster J, Harrison M, McEwan Paterson J, et al. (2013), The international linear collider technical design report - volume 1: Executive summary, arXiv 1306.6329.

64. ICFA. ICFA announces a new phase towards preparation for the International Linear Collider. ICFA (2020). Available at: https://www.interactions.org/press-release/icfa-announces-new-phase-towards-preparation-international.

65. Bernardi G, Brost E, Denisov D, Landsberg G, Aleksa M, d'Enterria D, et al. (2022), The future circular collider: A summary for the US 2021 Snowmass process, arXiv 2203.06520.

66. Agapov I, Benedikt M, Blondel A, Boscolo M, Brunner O, Chamizo Llatas M, et al. (2022), Future circular lepton collider FCC-ee: Overview and status, arXiv 2203.08310.

67. Benedikt M, Chance A, Dalena B, Denisov D, Giovannozzi M, Gutleber J, et al. (2022), Future circular hadron collider FCC-hh: Overview and status, arXiv 2203.07804.

68. CEPC Accelerator Study Group, (2019), arXiv 1901.03169. The CEPC input for the European Strategy for particle physics - accelerator.

69. CEPC-Accelerator-Study-Group, (2022), Snowmass2021 white paper AF3-CEPC, arXiv 2203.09451.

70. Tang J, Zhang Y, Xu Q, Gao J, Lou X, Wang Y, (2022), arXiv 2203.07987. Study overview for super proton-proton collider.

71. CLIC acceleratorAicheler M, Burrows PN, Catalan N, Corsini R, Draper M, Osborne J, et al. The compact linear collider (CLIC) - project implementation plan, 4 (2018). arXiv 1903.08655.

72. Brunner O, Burrows PN, Calatroni S, Catalan Lasheras N, Corsini R, D'Auria G, et al. (2022), The CLIC project, arXiv 2203.09186.

73. Stratakis D, Mokhov N, Palmer M, Pastrone N, Raubenheimer T, Rogers C, et al. (2022), A muon collider facility for physics discovery, arXiv 2203.08033.